# Identification of structural relaxation in the dielectric response of water

Jesper S. Hansen,[1] Alexander Kisliuk,[2] Alexei P. Sokolov,[2,3] and Catalin Gainaru[3,4,*]

[1] DNRF Centre "Glass and Time," IMFUFA, Department of Science and Environment, Roskilde University, Denmark
[2] Chemical Science Division, Oak Ridge National Laboratory, Oak Ridge, Tennessee 37831, United States
[3] Department of Chemistry, University of Tennessee, Knoxville, Tennessee 37996, United States
[4] Fakulty of Physics, Technical University of Dortmund, 44221 Dortmund, Germany

One century ago pioneering dielectric results obtained for water and n-alcohols triggered the advent of molecular rotation diffusion theory considered by Debye to describe the primary dielectric absorption in these liquids. Comparing dielectric, viscoelastic, and light scattering results we unambiguously demonstrate that the structural relaxation appears only as a high-frequency shoulder in the dielectric spectra of water. In contrast, the main dielectric peak is related to a supramolecular structure, analogous to the Debye-like peak observed in mono-alcohols.

Being in the focus of intensive research for the last few centuries [1], water still presents many challenging scientific puzzles. They include complex phase diagram [2], possible liquid-liquid transition [3] and significant role of quantum effects [4]. Among them is also the anomalously large dielectric constant that makes water an excellent solvent and is exploited on a daily basis in microwave heating. Not only the amplitude, but also the spectral shape of water's dielectric response is rather peculiar. For most liquids the dominating dielectric relaxation process is the structural α-relaxation that has asymmetric spectral shape corresponding to a stretched exponential relaxation in time domain [5,6]. In contrast, the dielectric spectrum in water is dominated by a Debye-like peak (single exponential process I), and has another less intense relaxation feature (process II) at higher frequencies [7,8]. The microscopic mechanism triggering this response remains highly debated with the focus on the main question: Does the Debye process reflect molecular scale structural relaxation or polarization of intrinsic supramolecular structures mediated by H-bonds?

In his seminal dielectric work [9] Debye himself argued in favor of the first scenario, based on hydrodynamic estimates of the rotational time for a single $H_2O$ molecule that appears close to the time scale of the process I, $\tau_I$. Several recent studies also assigned process I to reorientation diffusion of single water molecules [10,11,12]. In contrast, other phenomenological works consider process I related with dynamics of H-bonded network [13,14,15]. One major problem is that dielectric spectroscopy lacks microscopic information [5] and standalone cannot clarify the molecular nature of the processes observed for water. Hence information from other techniques needs to be involved.

In many aspects dielectric response of water resemble that known for mono-alcohols (MA) [16]. These liquids (e.g. n-propanol [17]), also display a bimodal dielectric spectra with dominating low-frequency Debye-like peak. Although Debye again assigned the main peak of n-propanol to rotational diffusion of single alcohol molecules [9], it is known now that this process has strikingly different microscopic origin [16,18]. The recent comparison of MA's characteristic dielectric times $\tau_I$ and $\tau_{II}$ with those reported from physical aging [19], NMR [20], calorimetric [21], viscoelastic [22], and light scattering [23] studies made possible the unambiguous identification of process II as the structural relaxation. The slow Debye process is currently assigned to dynamics of H-bonded networks in these systems. Confronting a widespread misperception [24], recent investigations performed on several H-bonded liquids revealed that this supramolecular process is not just a merely dielectric feature. It has been also identified in the depolarized light scattering (DLS) [25] and shear rheology [26] spectra, however, with a significantly lower intensity as compared with its dielectric counterpart.

Inspired by recent developments for MA, we pursue in this work the same strategy of combining dielectric, viscoelastic, and light scattering studies to unravel the nature of the dielectric processes in water. We accessed the viscoelastic signature of water's structural relaxation by means of computer simulations. Our results reveal that microscopic flow occurs in water on a time scale which is significantly shorter than $\tau_I$ but close to $\tau_{II}$, similar to MA. Our accurate DLS measurements discovered a low amplitude Debye-like relaxation process at frequencies below those characterizing structural relaxation [27,28,29]. These results provide unambiguous assignments of the dielectric processes in water.

For numerical studies of water we employed the SPC/Fw polarizable model that describes well many of its thermodynamic, structural, and kinetic properties [30]. During simulations the molecular pressure tensor is evaluated as:

$$\mathbf{P}(t) = \frac{1}{V}\left[ \sum_i m_i \mathbf{v}_i \mathbf{v}_i + \sum_i \sum_{j>i} \mathbf{F}_{ij} \mathbf{r}_{ij} \right], \qquad (1)$$

where $V$ is the system volume, $m$ the mass of water molecule, $\mathbf{v}$ is the center-of-mass velocity, $\mathbf{F}_{ij}$ is the force exerted by molecule $j$ on molecule $i$, and $\mathbf{r}_{ij} = \mathbf{r}_i - \mathbf{r}_j$, with $\mathbf{r}$ the center-of-mass position vector. $\mathbf{v}_i\mathbf{v}_i$ and $\mathbf{F}_{ij}\mathbf{r}_{ij}$ are outer vector products generating a second order tensor. The symmetric part of the pressure tensor was extracted as $\mathbf{P_s} = (\mathbf{P}+\mathbf{P}^T)/2$. The stress autocorrelation function, calculated using the off-diagonal elements of the symmetric pressure tensor

$$C(t) = \frac{V}{3k_BT}\sum_{\alpha\beta}\langle \mathbf{P}_{s,\alpha\beta}(t)\mathbf{P}_{s,\alpha\beta}(0)\rangle, \quad (2)$$

(here indices $\alpha\beta$ run over the off-diagonal tensor elements $xy$, $xz$, $yz$) can be identified with the shear modulus relaxation function via the fluctuation-dissipation theorem. Frequency-dependent viscosity $\eta^*(\omega)$ was evaluated using the one-sided Fourier transform of Eq. (2).

The values for the steady-state viscosity estimated as $\eta_0 = \lim_{\omega\to 0}\mathrm{Re}[\eta^*(\omega)]$ are listed in Table 1 for the different temperatures considered in the present study. The good agreement between $\eta_0$ and experimental literature data $\eta_{0,\mathrm{Exp}}$ [31] (also included in Table 1) brings confidence to our chosen approach.

Table 1. Rheological parameters of water

| $T$ (K) | $\eta_0$ (mPa·s) | $\eta_{0,\mathrm{Exp}}$ (mPas) | $\tau_s$ (ps) | $G_\infty$ (GPa) |
|---|---|---|---|---|
| 319 | 0.61 | 0.57 | 0.22 | 2.8 |
| 309 | 0.67 | 0.69 | 0.27 | 2.5 |
| 299 | 0.97 | 0.85 | 0.33 | 2.9 |
| 289 | 1.07 | 1.09 | 0.39 | 2.7 |
| 284 | 1.42 | 1.23 | 0.46 | 3.1 |
| 278 | 1.55 | 1.40 | 0.5 | 3.1 |

At short times the simulated stress autocorrelation functions (Fig. 1) are dominated by vibrational contributions which are practically $T$-invariant. By lowering the temperature the long-time decay which corresponds to structural relaxation progressively slows down. In the relatively small dynamic range covered by our simulations the shear autocorrelation functions, although plotted on double-logarithmic scales, do not reveal the presence of two relaxations. As demonstrated in the inset of Fig. 1, $T$-dependent horizontal shifts collapse all datasets to a master curve, demonstrating the applicability of time-temperature superposition for the main relaxation process of water.

The transition from elastic to viscous regime can be described reasonably well by a Kohlraush function $C(t) \propto \exp[-(t/\tau_K)^{\beta_K}]$ with stretching exponent $\beta_K \approx 0.73$. Using the parameter $\tau_K$ characterizing the master curve (inset Fig. 1) and the $T$-dependent scaling factors we estimated the values of shear relaxation times $\tau_s$ which are included in Table 1 and plotted in Fig 3(a).

The DLS spectrum of water was measured at 298 K using a Raman spectrometer and a Fabry-Perot interferometer, as previously done in [27]. The experiments were performed in backscattering geometry using laser wavelength 532 nm, and a power of 100 mW at the sample position. For achieving a good accuracy of the spectrum at low frequencies the accumulation time was extended to over 48 hours.

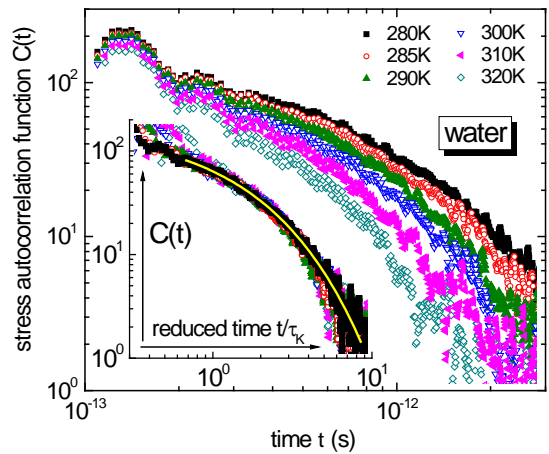

Fig. 1 (Color online) Temperature evolution of stress correlation function C(t) obtained from computer simulations. The inset shows the master curve obtained by the horizontal shift of the individual C(t) datasets. The (yellow) solid line is a fit with a stretched exponential function.

The DLS susceptibility $\chi''_{\mathrm{DLS}}$ spectrum of water is compared in Fig. 2 with its dielectric counterpart as previously published in [29]. From the high frequency side the DLS spectrum is dominated by the vibrational band followed at intermediate ν (of about 200 GHz) by the α-relaxation. The spectral shape of the current measurements are in good agreement with previous works [27,29], with one significant exception: in our data a shoulder at about 30-50 GHz reveals the existence of a process which is slower than structural α-relaxation.

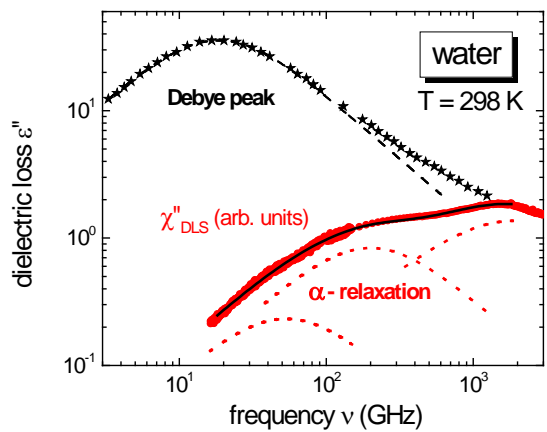

Fig. 2 (Color online) Comparison of dielectric (black stars, taken from Ref. [29]) and DSL (red dots, current work) susceptibilities for water at room temperature. The dashed black line corresponds to a Debye function. The solid black line is a fit of DLS spectrum with the sum of two Debye functions, accounting for the contributions of relaxation processes, and an arbitrary peak function considered for the fast dynamics. The red dotted lines highlight the individual contributions composing the DLS spectrum.

At first glance this observation contradicts previous statement that "a single relaxation…is sufficient" to describe the slow DLS dynamics of water [29]. However, a close inspection of data plotted in Fig. 2 of Ref. [29] reveals that such single-relaxation approach fails to describe not only present, but also previously published data for frequencies below 100 GHz. The improved accuracy of our measurements demonstrates that a good description of the *entire* DLS spectrum requires *two* relaxations peaks in addition to the vibration dynamics. It is obvious that the position of the α-peak in the DLS spectrum (Fig. 2) corresponds to a frequency range at which significant deviations from the Debye behavior occur in the dielectric response.

The characteristic times obtained in the present study are plotted together with literature data in Fig. 3(a). To demonstrate the high similarities between the relaxation map of water and that of the archetypical MA n-propanol, we included for the latter recently published dielectric, rheological, and DLS characteristic times in Fig. 3(b).

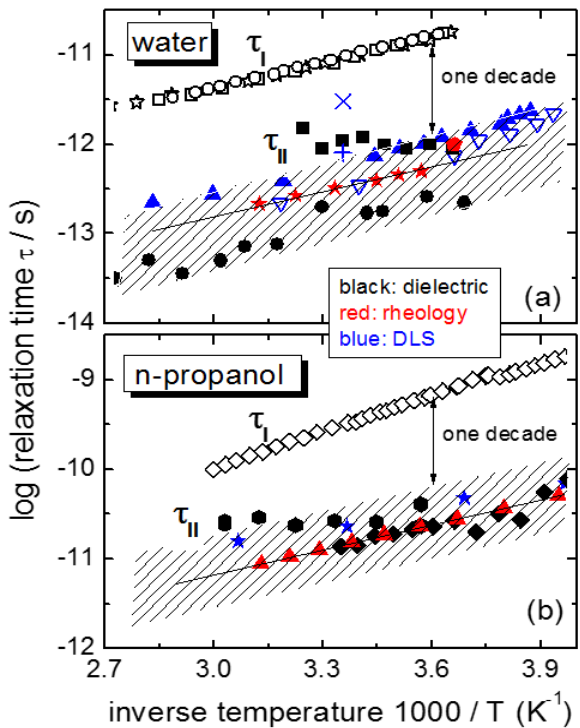

Fig. 3 (Color online) Compilation of dielectric (black symbols), viscoelastic (red symbols), and light scattering (blue symbols) time constants of (a) water and (b) n-propanol. Literature include in (a) for dielectrics [10] (open circles), [8] (open stars and filled circles), and [13] (open circles and filled squares), for shear rheology [34] (filled pentagons) and for light scattering [27] (filled triangles) and [28] (open triangles). In (b) the dielectric data are taken from [18] (open and filled diamonds), and [23] (filled hexagons), while rheology and light scattering data are also from [23]. The rheology and DLS time constants obtained in this work are plotted in (a) as filled stars and plus (for α-process) and cross (for the slower mode) symbols, respectively. The two parallel solid lines are fits with Arrhenius laws.

Focusing first on the dielectric results, one could easily observe the excellent agreement between the various sources [8,10,13] regarding the time constants of the Debye peak. On the other hand, the fact that the secondary dielectric process in both liquids strongly overlaps with the dominant Debye contribution renders the identification of the characteristic time $\tau_{II}$ as model dependent. As a result, a range (marked by the dashed areas in Fig. 3) of values is reported for $\tau_{II}$. Nonetheless, these values are smaller than $\tau_I$ by at least one decade at T≈278 K for both liquids (Fig. 3). Since for MA [16], including n-propanol [18], $\tau_{II}$ was previously ascribed to the structural relaxation time, the high similarities between Fig. 3(a) and Fig. 3(b) suggest that this assignment may also hold for water. Light scattering data strongly supports this conclusion, as demonstrated by $\tau_\alpha$ values which are lying near the $\tau_{II}$ values reported in [13] (Fig. 3a). We note here that the $\tau_\alpha$ results from DLS and implicitly dielectric $\tau_{II}$ are in agreement with previous X-ray scattering [32] and NMR [33] relaxation studies of water.

Let us now focus on the rheology results obtained from computer simulations. The $\tau_s$ values are comparable with the previous rheological investigations and with the structural relaxation time estimated from DLS (Fig. 3a). For both liquids the temperature evolution of $\tau_s$ is similar to that exhibited by the other time constants contained in the dashed areas of Figs. 3. Moreover, in this temperature range both $\tau_s(T)$ datasets can be described by Arrhenius laws $\tau_s \propto \exp(E/k_B T)$ (represented as solid lines in Fig. 3) with a common activation energy $E$ of about 15 kJ/mol, which suggests a similar underlying mechanisms for the structural relaxation process of both water and n-propanol.

From a quantitative point of view, the current viscoelastic results reveal the following inconsistency which has been largely overlooked: taking into account that at room temperature the shear viscosity of water is about $10^{-3}$ Pas, and assuming that the Debye process is a structural relaxation with $\tau_\alpha \sim \tau_I = 8.4$ ps [29], Maxwell relation predicts for the instantaneous shear modulus $G_\infty = \eta_0/\tau_\alpha \sim 10^8$ Pa, a value which is too small to be considered as realistic. On the other hand, considering $\tau_\alpha = \tau_s$ *i.e.* at least ten times smaller than $\tau_I$, the $G_\infty$ values calculated using Maxwell relation become on the order of GPa (see Table 1), in good agreement with experimental predictions [34]. All these results clearly support assignment of the high frequency dielectric process II to structural relaxation.

Not only the dynamic (Figs. 3), but also the static dielectric behavior is similar for water and n-alcohols. The strength of Debye process is too large, while that of structural relaxation is too small as compared with theoretical expectations for molecular dipoles lacking orientational correlations. For MA the primary response is assigned to fluctuations of the collective dipole accumulated along the contour of quasilinear supramolecular structures [20]. For water that has a tetrahedral structure the significant degree of static correlations established within the first molecular coordination shell is usually invoked to explain its large dielectric absorption [35]. Regarding α-dynamics, for MA it was demonstrated to be related to reorientation of alkyl units around the backbone formed by H-bonded hydroxyl groups, hence it involves only a fraction of the total molecular dipole moment [20]. For water the activation energy of process II which corresponds to breaking of a single H-bond, in harmony with previous computer simulation studies [12], also suggests a restricted reorientation of molecular dipoles, since an isotropic dynamics would involve the breaking of at least two H-bonds (close to room temperature the number of H-bonds per water molecule varies between 2 and 4 [36]).

Regarding the slow water dynamics, the newly identified terminal DLS relaxation is faster than the dielectric Debye process by a factor close to 3. This ratio is known for many processes and is well justified by the difference between vectorial and tensorial characters of the responses probed by dielectric spectroscopy and DLS, respectively [37]. In this respect the low-frequency process appears to be governed by a rotational diffusive mechanisms, while α-process involves a significant amount of large-angle reorientations, as also suggested by previous studies [12]. Another argument for a common origin of the dielectric and DLS slow processes is the amplitude of the latter. Its small DLS intensity reflects a low optical polarizability usual for processes with large dielectric response, as the one associated here with the Debye process.

Having clarified the nature of water's dielectric processes, we want to discuss why the Debye-Stokes relation $\eta_0 = k_B T \tau_\alpha/(4\pi R^3)$ with *geometrical* radius $R$ of water molecule provides $\tau_\alpha \approx \tau_I$. First of all, classical hydrodynamic approaches do not hold on molecular level. It is well established that for ordinary liquids Debye-Stokes and Stokes-Einstein relations yield a radius of reorienting/translating moieties $R_H$ much smaller than molecular $R$ [38]. Applying for water the empiric relation $R_H \approx R/2$ [38], Debye-Stokes relation will predict a value for $\tau_\alpha \sim \tau_I/8 \approx \tau_{II}$ consistent with the secondary dielectric process [13].

In conclusion, the presented detailed experimental and computational studies, their analysis and discussion of literature data provide unambiguous assignment of the high frequency dielectric process to the structural α-relaxation of water. The intense low-frequency Debye-like peak in water is a supramolecular process, analogous to the Debye-like process known for monoalcohols. In other words, microwave heating operating at 2.45 GHz should not be directly connected with the reorientation process of single water molecules. From the general perspective emerged from recent studies of other H-bonded liquids, the dielectric Debye process is the manifestation of polarization fluctuations associated with supramolecular (tetrahedral) structure of water. The current DLS results open the venue for future investigations of this puzzling process by other techniques which are usually employed to complement dielectric spectroscopy. Our viscoelastic investigations covering a narrow dynamic range did not reveal the presence of such slow process. However, recently gained knowledge in MA's behavior suggests that ultrafast rheology [39] might also detect in near future the signature of supramolecular dynamics which governs the anomalous dielectric behavior of water.

We thank Prof. Roland Böhmer and Dr. Tina Hecksher for interesting discussions. JSH thanks Lundbeckfonden for supporting this work as a part of grant no. R49-A5634. US participants greatly acknowledge partial financial support from NSF Polymer program (DMR-1408811).

———————————

*Corresponding author.
Electronic address: catalin.gainaru@uni-dortmund.de